\documentclass[12pt]{article}

\begin{document}
\begin{center}
{\bf On superluminal fermions within the second derivative equation}\\
\vspace{5mm}
 S. I. Kruglov \\
\vspace{5mm}
\textit{Department of Chemical and Physical Sciences, University of Toronto,\\
3359 Mississauga Rd. North, Mississauga, Ontario, Canada L5L 1C6}
\end{center}

\begin{abstract}
We postulate the second-order derivative equation with four parameters for spin-1/2 fermions possessing
two mass states. For some choice of parameters fermions propagate with the superluminal speed. Thus, the novel tachyonic equation is suggested. The relativistic 20-component first-order wave equation is formulated and projection operators extracting states with definite energy and spin projections are obtained. The Lagrangian formulation of the first-order equation is presented and the electric current and energy-momentum tensor are found. The minimal and non-minimal electromagnetic interactions of fermions are considered and Schr\"{o}dinger's form of the equation and the quantum-mechanical Hamiltonian are obtained. The canonical quantization of the field in the first-order formalism is performed and we find the vacuum expectation of chronological pairing of operators.

\end{abstract}

\section{Introduction}

The study of models which allow the superluminal behavior of particles is of definite theoretical interest. Previously, models of superluminal particles (tachyons) were considered in \cite{Chodos}, \cite{Recami} (see refs. therein).
Here, we pay attention on the possibility of the superluminal propagation of fermions in the framework of the higher derivative (HD) theory. HD models attract attention due to their improved renormalization properties \cite{Thirring}, \cite{Pais}, \cite{Heisenberg}, \cite{Lee}, \cite{Hawking}. We formulate here the second-order (in derivatives) equation for spin-1/2 fields, which admits the superluminal particle speed and is the generalization of Barut's equation \cite{Barut}, \cite{Barut1} describing fermions with two mass states.

The paper is organized as follows. In Sec.2, we formulate the second derivative equation
for free fermions with the dispersion relation allowing the superluminal propagation.
The first-order relativistic wave equation is derived and
the relativistically invariant bilinear form is obtained in Sec.3. The mass and spin projection operators are given in Sec.4. We obtain the electric current and energy-momentum tensor in Sec.5.
Sec.6 is devoted to the introduction of minimal and non-minimal electromagnetic interactions. The Schr\"{o}dinger form of the equation is given and quantum-mechanical Hamiltonian is obtained in Sec.7.  We consider the canonical quantization of a model and obtain the the vacuum expectation of chronological pairing of operators in Sec.8. In Sec.9 the novel tachyonic equation is suggested. The results are discussed in Sec.10. Appendices A and B  contain some useful products of matrices. We use Euclidian metrics and notations as in \cite{Ahieser}, and the system of units $\hbar =c=1$ is chosen. Greek letters run 1,2,3,4, and Latin letters run 1,2,3.

\section{The Model}

Let us investigate the second-order (in derivatives) field equation with four parameters
describing spin-1/2 particles. We suggest HD equation
\begin{equation}
\left(\gamma_\mu\partial_\mu-\frac{a}{M}\partial_\mu^2+M+b\gamma_\mu\gamma_5\partial_\mu+M_0 \gamma_5\right)\psi(x)=0,  \label{1}
\end{equation}
where $\partial_\nu =\partial/\partial x_\nu =(\partial/\partial
x_m,\partial/\partial (it))$, $\psi (x)$ is a Dirac spinor and the
Dirac matrices $\gamma_\mu $ obey the commutation relations
$\gamma_\mu \gamma_\nu +\gamma_\nu \gamma_\mu =2\delta_{\mu\nu}$, $\gamma_5=\gamma_1\gamma_2\gamma_3\gamma_4$.
The parameters $a$ and $b$ are dimensionless, and $M$, $M_0$ have the dimension of the mass. The suggested equation violates in general the parity conservation.
Equation (1) is the generalization of the Barut's equation \cite{Barut}, \cite{Barut1}. At $b=0$, $M_0=0$, one comes to the Barut equation which was interpreted as the equation for a description of $e$ and $\mu$ leptons (see also \cite{Kruglov}). One can obtain from Eq.(1) the tachyonic first-order Dirac equation by putting $a=0$, $b=0$, $M=0$ ($a/M=0$).
We pay also attention here on the particular case of Eq.(1) when parameters $a$, $b$, $M$ and $M_0$ are connected with three mass parameters $\kappa_1$, $\kappa_2$, $\kappa_3$ as follows:
\begin{equation}
M=\frac{\kappa_1\kappa_2}{\kappa_1+\kappa_2},~~~a=-\frac{M}{\kappa_1+\kappa_2},~~~
b=\frac{\kappa_3}{\kappa_1+\kappa_2},~~~M_0=b\kappa_1.
 \label{2}
\end{equation}
At this case Eq.(1) is reduced to the equation
\begin{equation}
\left(\gamma_\nu\partial_\nu+ \kappa_1\right)
\left(\gamma_\mu\partial_\mu+ \kappa_2+\kappa_3\gamma_5\right)\psi(x)=0 . \label{3}
\end{equation}
At $\kappa_3=0$, one arrives at Barut's equation \cite{Barut}, \cite{Barut1}. In general, four parameters $a$, $b$, $M$, and $M_0$ in Eq.(1) are independent. In momentum space Eq.(3) becomes
\begin{equation}
\left(i\hat{p}+ \kappa_1\right)
\left(i\hat{p}+ \kappa_2+\kappa_3\gamma_5\right)\psi(p)=0 . \label{4}
\end{equation}
where $\hat{p}=p_\mu\gamma_\mu$, $p_\mu=(\textbf{p},ip_0)$.
After multiplying Eq.(4) by the matrix
$\left(i\hat{p}- \kappa_2+\kappa_3\gamma_5\right)\left(i\hat{p}- \kappa_1\right)$,
one obtains the dispersion equation
\begin{equation}
\left(p^2+ \kappa_1^2\right)
\left(p^2+ \kappa_2^2-\kappa_3^2\right)=0,  \label{5}
\end{equation}
where $p^2=\textbf{p}^2-p_0^2$, and $p_0$ is the energy of the fermion.
It follows from Eq.(5) that fermions possess two mass states with masses $\kappa_1$ and $\kappa=\sqrt{\kappa_2^2-\kappa_3^2}$. At $\kappa_2>\kappa_3$ fermions have real masses, but at $\kappa_2<\kappa_3$ the mass $\kappa$ becomes imaginary that indicates on the presence of a tachyon. In this case ($\kappa_2<\kappa_3$) fermions propagate with the superluminal speed. If the mass $\kappa_1$ is huge, such state of fermions is not observable. One can speculate that such state of fields with big mass $\kappa_1$ and weakly interacting with fields contribute to dark matter. In the case $\kappa_2=\kappa_3\neq 0$, we have massless state. Another possible way to realize a massless state is to choose $\kappa_2=\kappa_3=0$. If $\kappa_3=i\varphi$, where $\varphi$ is a real number, the second state is not tachyonic and corresponds to the mass $\kappa=\sqrt{\kappa_2^2+\varphi^2}$. Such situation is realized in the the Nambu$-$Jona-Lasinio model with CP-violating condensate $\varphi=i\langle \bar{\psi}\gamma_5\psi\rangle\neq 0$ \cite{Kr}.
Thus, introducing three mass parameters allow us to investigate differing mass states: massive, massless and tachyonic. Propagator corresponding to the model based on Eq.(4) is given by
\[
\left[\left(i\hat{p}+ \kappa_1\right)
\left(i\hat{p}+ \kappa_2+\kappa_3\gamma_5\right)\right]^{-1}
\]
\vspace{-7mm}
\begin{equation} \label{6}
\end{equation}
\vspace{-7mm}
\[
=\frac{\left(i\hat{p}- \kappa_2+\kappa_3\gamma_5\right)\left(i\hat{p}- \kappa_1\right)}
{\kappa_1^2-\kappa^2}\left[\frac{1}{p^2+\kappa^2}-\frac{1}{p^2+\kappa_1^2}\right].
\]
where $\kappa^2=\kappa_2^2-\kappa_3^2$. Thus, we have the difference of propagators in Eq.(6) so that the state with the mass $\kappa_1$ is a ghost. Another possibility is to interpret the state with the mass $\kappa$ as a ghost. These two cases correspond to different signs in the Lagrangian. We interpret here the state with the mass $\kappa_1$ as a ghost state. When the mass $\kappa_1\rightarrow\infty$ that state is not observable. But ultraviolet behavior of the propagator with the mass $\kappa$ is improved by this additional subtraction. The situation is similar to the Pauli$-$Villars regularization. The difference is that the state with the mass $\kappa_1$ is in the spectrum. It should be noted that similar scheme with higher derivative equations was explored in the Lee$-$Wick model \cite{Lee}.
In the general case, when four parameters $a$, $b$, $M$, and $M_0$ are independent, the dispersion equation corresponding to Eq.(1) reads
\begin{equation}
a^2p^4+M^2\left(1+2a-b^2\right)p^2+M^2\left(M^2-M^2_0\right)=0.
\label{7}
\end{equation}
This equation is valid for a particular case $M_0=0$, $b\neq 0$ neglecting Eqs.(2). Then finding the roots of Eq.(7) at $M_0=0$, one obtains squared of masses of the field states
\[
m_1^2=\frac{M^2}{2a^2}\left[1+2a-b^2 +\sqrt{\left(1+4a-b^2\right)\left(1-b^2\right)}\right],
\]
\vspace{-7mm}
\begin{equation} \label{8}
\end{equation}
\vspace{-7mm}
\[
m_2^2=\frac{M^2}{2a^2}\left[1+2a-b^2 -\sqrt{\left(1+4a-b^2\right)\left(1-b^2\right)}\right].
\]
If $b=\pm 1$, one has $m_1^2=m_2^2$ and for $a>0$ fields have subluminal speed but for $a<0$ fields propagate with superluminal speed and are tachyons. At $b=0$, $a\neq 0$, we arrive at the case considered in \cite{Kruglov}. If $|b|<1$ and $1+2a-b^2>0$ fields have subluminal speed but for $1+2a-b^2<0$ fields propagate with superluminal speed. Thus, Eq.(1) allows us to consider different models of superluminal particles.

\section{First-Order Field Equation}

Let us introduce 20-component function which is the direct sum of a bispinor $\psi (x)$ and a vector-bispinor $\psi_\mu (x)$:
\begin{equation}
\Psi (x)=\left\{ \psi _A(x)\right\} =\left(
\begin{array}{c}
\psi (x)\\
\psi _\mu (x)
\end{array}
\right) , \label{9}
\end{equation}
where the index $A$ runs values $A=0,\mu$. The vector-bispinor is given by
\begin{equation}
  \psi_\mu (x)=-\frac{1}{M}\partial_\mu \psi (x).
\label{10}
\end{equation}
Thus, the wave function $\Psi(x)$ presents the reducible representation of the Lorentz group.
Now, we use the elements of the entire matrix
algebra $\varepsilon ^{A,B}$ \cite{Kruglov1} with matrix elements and the product of two matrices:
\begin{equation}
\left( \varepsilon ^{M,N}\right) _{AB}=\delta _{MA}\delta _{N,B},
\hspace{0.5in}\varepsilon ^{M,A}\varepsilon ^{B,N}=\delta
_{AB}\varepsilon ^{M,N}, \label{11}
\end{equation}
where indexes $A,B,M,N$ run five values $0,1,2,3,4$. Any matrix with matrix elements $a_{ij}$ can be written as $a_{ij}\varepsilon ^{i,j}$, where we imply a summation on repeated indexes. Then Eq.(1), taking into account Eqs.(9)-(11), becomes
\[
\partial _\nu\left(\varepsilon ^{\nu,0 }+ a\varepsilon ^{0,\nu}+
\varepsilon^{0,0}\gamma_\nu \left(1+b\gamma_5\right) \right)_{AB}\Psi_{B}(x)
\]
\vspace{-7mm}
\begin{equation} \label{12}
\end{equation}
\vspace{-7mm}
\[
+\left[M\left(\varepsilon ^{0,0 }+ \varepsilon ^{\mu,\mu}\right)+M_0\varepsilon ^{0,0 }\gamma_5\right]_{AB} \Psi_{B} (x)=0 .
\]
Equation (12) can be cast into the first-order relativistic wave equation
\begin{equation}
\left( \Gamma _\nu \partial _\nu +M +M_0\Gamma_5\right) \Psi (x)=0 , \label{13}
\end{equation}
where 20-component matrices are given as follows:
\begin{equation}
\Gamma_\nu =\left(\varepsilon ^{\nu,0 }+ a\varepsilon ^{0,\nu
}\right)\otimes I_4 + \varepsilon^{0,0}\otimes\gamma_\nu\left(1+b\gamma_5\right),
~~~~\Gamma_5=\varepsilon^{0,0}\otimes\gamma_5,
\label{14}
\end{equation}
and $I_4$ is the unit $4\times 4$ matrix acting in the bispinor subspace,  $\otimes$ is
the direct product of matrices. At $b=0$, $M_0=0$ Eq.(13) is converted into equation considered in \cite{Kruglov}. To investigate the properties of first-order relativistic wave equation (13), one can use the general methods described in \cite{Gel'fand}. Eq.(13) is convenient for different
applications as the matrices of the equation, $\Gamma _\nu$, $\Gamma _5$, are
expressed through the elements of the entire matrix algebra (11).

The generators of the Lorentz group in the representation of the wave function (9)
\[
\left[(1/2,0)\oplus (0,1/2)\right]\oplus\left\{(1/2,1/2)\otimes
\left[(1/2,0)\oplus (0,1/2)\right]\right\}
\]
are given by \cite{Kruglov}
\begin{equation}
J_{\mu \nu }=\left(\varepsilon^{\mu,\nu}-\varepsilon^{\nu,\mu}\right)\otimes I_4+I_5 \otimes
\frac{1}{4}\left( \gamma_\mu
\gamma_\nu-\gamma_\nu \gamma_\mu \right) , \label{15}
\end{equation}
and obey the commutation relations:
\begin{equation}
\left[ J_{\mu \nu },J_{\alpha \beta}\right] =\delta
_{\nu \alpha }J_{\mu \beta}+\delta _{\mu \beta }J_{\nu
\alpha}-\delta _{\nu \beta }J_{\mu \alpha}-\delta
_{\mu \alpha }J_{\nu \beta} . \label{16}
\end{equation}
The relativistic wave equation (13) is form-invariant under the Lorentz transformations. Indeed, one can verify that matrices (14) obey the necessary commutation equations
\begin{equation}
\left[\Gamma_\lambda , J_{\mu \nu }\right] =\delta
_{\lambda\mu }\Gamma_{\nu }-\delta _{\lambda\nu}\Gamma_{\mu},~~~~
\left[\Gamma_5,J_{\mu \nu }\right]=0. \label{17}
\end{equation}
The relativistically invariant bilinear form is
\begin{equation}
\overline{\Psi }\Psi =\Psi ^{+}\eta \Psi ,  \label{18}
\end{equation}
where $\Psi ^{+}$ is the Hermitian-conjugate wave function and the Hermitianizing matrix $\eta$
is given by \cite{Kruglov}
\begin{equation}
\eta=\left(a\varepsilon^{m,m}-a\varepsilon^{4,4}-\varepsilon^{0,0}\right)\otimes
\gamma_4 , \label{19}
\end{equation}
$\eta^+=\eta$, and obeys equations:
\begin{equation}
\eta \Gamma_i=-\Gamma _i^{+}\eta,~~~~\eta \Gamma _4=\Gamma
_4^{+}\eta,~~~~\eta \Gamma _5=-\Gamma
_5\eta ~~~~(i=1,2,3) .  \label{20}
\end{equation}
It should be noted that the Lorentz-invariant $\overline{\Psi}\Gamma_5\Psi$ is imaginary, $(\overline{\Psi}\Gamma_5\Psi)^+ =-(\overline{\Psi}\Gamma_5\Psi)$. Therefore, to formulate the variational principle, one has to imply that the parameter $M_0$ is imaginary, $M_0=iM_1$, where $M_1$ is a real parameter.
Then the ``conjugated" equation reads
\begin{equation}
\overline{\Psi }(x)\left(\Gamma _\mu \overleftarrow{\partial}_\mu
-M-iM_1 \Gamma_5\right)=0 .
\label{21}
\end{equation}
Eq.(21) follows from Eq.(13) (at $M_0=iM_1$) by Hermitian conjugation and
multiplying the equation by $\eta$.

\section{ Mass and Spin Projection Operators}

Here, we restrict our consideration by putting $M_0=0$ ($M_1=0$); then masses of fermionic fields
$m_1$ and $m_2$ are given by Eq.(8). Because there are two field states, one can introduce an
additional quantum number $\tau$ in such a way that $\tau=1$, $2$ for states with masses $m_1$ and $m_2$, correspondingly \cite{Wilson}. Let us consider states of particles with definite energy, $p_0$,
and momentum, $\textbf{p}$. For the definite mass state, we have
\begin{equation}
p_0=\sqrt{\textbf{p}^2 +m_1^2},~~~ or ~~~ p_0=\sqrt{\textbf{p}^2
+m_2^2}. \label{22}
\end{equation}
The index $\tau$ in four momentum $p=(\textbf{p},ip_0)$ is omitted here. Solutions to Eq.(13)
with definite energy-momentum in the form of plane waves are
given as follows:
\begin{equation}
\Psi^{(\pm)}_{\pm p}(x)=\frac{1}{\sqrt{2p_0 V}}U(\pm p)\exp(\pm
ipx) , \label{23}
\end{equation}
where $V$ is the normalization volume. Substituting Eq.(23) into Eg.(13),
we obtain (at $M_0=0$)
\begin{equation}
\left(\pm i \check{p}+ M \right)U(\pm p)=0 , \label{24}
\end{equation}
where $\check{p}=\Gamma_\mu p_\mu$, and the solution $\Psi^{(+)}_{+
p}(x)$ corresponding to the positive energy $p_0$ describes particles
and the solution $\Psi^{(-)}_{- p}(x)$ with negative energy
$-p_0$ corresponds to antiparticles. Similar to QED \cite{Ahieser}, we use here the
normalization condition
\begin{equation}
\int_V \overline{\Psi}^{(\pm)}_{\pm p}(x)\Gamma_4
\Psi^{(\pm)}_{\pm p}(x)d^3 x=1 , \label{25}
\end{equation}
where $\overline{\Psi}^{(\pm)}_{\pm p}(x)=\left(\Psi^{(\pm)}_{\pm
p}(x)\right)^+ \eta$, and the integration over the
volume $V$ is implied in Eq.(25). Eqs.(23)-(25) are valid for two
values of energy-momentum corresponding to additional quantum number $\tau =1,2$.
The 20-component functions $U(\pm p)$ obey equations:
\begin{equation}
\overline{U}(\pm p)\Gamma_\mu U(\pm p)=-2ip_\mu , \label{26}
\end{equation}
\begin{equation}
\overline{U}(\pm p)U(\pm p)=\mp 2\frac{p^2}{M}, \label{27}
\end{equation}
where $\tau=1$ corresponds to $p^2=-m_1^2$, and  $\tau=2$ corresponds to $p^2=-m_2^2$.
Eqs.(26),(27) are similar to normalization conditions for bispinors
in QED. To obtain the projection matrix extracting solutions to Eq.(24), we
use the minimal equation for the matrix $\check{p}=p_\mu \Gamma_\mu$ (see Appendix A):
\begin{equation}
\check{p}^5 -\left(1+2a-b^2\right)p^2 \check{p}^3 + a^2 p^4
\check{p}=0 . \label{28}
\end{equation}
With the help of the method of the work \cite{Fedorov}, one obtains from Eq.(28)
the projection matrices
\begin{equation}
\Pi_{\pm}=\frac{\pm i\check{p}\left(M\mp i\check{p}\right)
\left[\check{p}^2-\left(1+2a-b^2\right)p^2-M^2\right]}{2M^2
\left[\left(1+2a-b^2\right)p^2+2M^2\right]}, \label{29}
\end{equation}
that obey equations
\begin{equation}
\left(\pm i \check{p}+ M \right)\Pi_{\pm}=0 , \label{30}
\end{equation}
\begin{equation}
\Pi_{\pm}^2=\Pi_{\pm} ,~~~~\Pi_{+}\Pi_{-}=0 .\label{31}
\end{equation}
The operator of the spin projections on the direction of the
momentum $\textbf{p}$ is given by \cite{Kruglov}
\begin{equation}
\sigma_p=-\frac{i}{2|\textbf{p}|}\epsilon_{abc}\textbf{p}_a J_{bc}
, \label{32}
\end{equation}
where $|\textbf{p}| =\sqrt{p_1^2 +p_2^2+p_3^2}$, and obeys the matrix equation
\begin{equation}
\left(\sigma_p^2-\frac{1}{4}\right)\left(\sigma_p^2-\frac{9}{4}\right)=0.
\label{33}
\end{equation}
From Eq.(33), we find the spin projection operator \cite{Kruglov}
\begin{equation}
P_{\pm 1/2}=\mp\frac{1}{2}\left(\sigma_p\pm\frac{1}{2}
\right)\left(\sigma_p^2-\frac{9}{4}\right) \label{34}
\end{equation}
which obeys the equation
\begin{equation}
\sigma_p P_{\pm 1/2}=\pm \frac{1}{2} P_{\pm 1/2} . \label{35}
\end{equation}
One can verify that the operators $\check{p}$ and $\sigma_p$
commute $[ \check{p},\sigma_p]=0$ and, therefore, they
have the common eigenfunctions.
Using Eqs.(30),(35  ), one obtains the projection
operators for pure spin states in the form of matrix-dyads
\begin{equation}
U_s(\pm p) \bullet \overline{U}_s(\pm p)=N_{\pm} P_{\pm 1/2}\Pi_\pm ,
\label{36}
\end{equation}
with matrix elements $\left(U_s(\pm p) \bullet
\overline{U}_s(\pm p)\right)_{AB}=\left(U_s(\pm p)\right)_A
\left(\overline{U}_s(\pm p)\right)_B$, and $N_{\pm}$ is the normalization constant.
We have introduced the spin index $s=\pm 1/2$ in Eq.(36). Eq.(36) can be used for calculations of different quantum processes
of particles obeying Eq.(1) (for $M_0=0$) in the first-order formalism.
We use the normalization for pure spin states as follows:
\begin{equation}
\overline{U}_s(\pm p)U_r(\pm p)=\mp \delta_{sr}\frac{p^2}{M}
,~~~~\overline{U}_s( p)U_r(-p)=0 .\label{37}
\end{equation}
Eq. (27) follows from Eq.(37) after a summation over the spin index $s=\pm 1/2$.
Taking into consideration the relationship (see
Appendix A)
\begin{equation}
 \check{p}\left( P_{1/2}+ P_{- 1/2}\right)=\check{p},
\label{38}
\end{equation}
one can obtain from Eqs.(36),(37) an expression for matrix density for impure spin states
\begin{equation}
\sum_s U_s(\pm p) \bullet \overline{U}_s(\pm p)=N_{\pm} \Pi_{\pm } .
\label{39}
\end{equation}
Using the equation tr$\Pi_{\pm }=2$ (see Appendix A) and taking the trace in both sides of Eq.(39),
we find
\begin{equation}
 \sum_s \overline{U}_s(\pm p)U_s(\pm p)=2N_{\pm} .
\label{40}
\end{equation}
Comparing Eq.(40) with Eq.(27), one obtains
\begin{equation}
 N_{\pm}=\mp \frac{p^2}{M}.
\label{41}
\end{equation}
Eqs.(29),(39) generalize expressions obtained in \cite{Kruglov2} on the case $b\neq 0$.

\section{Electric Current and Energy-Momentum Tensor}

Now, we consider the case $M_0=iM_1$; then Eqs.(13),(21) follow from
the Lagrangian
\[
{\cal L}=-\frac{1}{2}\biggl[\overline{\Psi }(x)\left(\Gamma _\nu \partial _\nu
+M +iM_1\Gamma_5\right)\Psi (x)
\]
\vspace{-7mm}
\begin{equation} \label{42}
\end{equation}
\vspace{-7mm}
\[
-\overline{\Psi }(x)\left(\Gamma _\mu \overleftarrow{\partial}_\mu
-M-iM_1 \Gamma_5\right)\Psi (x)\biggr],
\]
which is a real function. The electric current density is given by \cite{Bogolyubov}
\begin{equation}
j_\mu (x)=i\left(\overline{\Psi }(x)\frac{\partial {\cal L}}{\partial\left(\partial_\mu\overline{\Psi }(x)\right)}- \frac{\partial {\cal L}}{\partial\left(\partial_\mu\Psi (x)\right)}\Psi(x)\right).
\label{43}
\end{equation}
From Eqs.(42),(43),(9),(11), we obtain
\[
j_\mu (x)=i\overline{\Psi }(x)\Gamma_\mu \Psi(x)=
\]
\vspace{-7mm}
\begin{equation} \label{44}
\end{equation}
\vspace{-7mm}
\[
= -i\overline{\psi}(x)\gamma_\mu\left(1+b\gamma_5\right) \psi(x)+
\frac{ia}{M}\left[\overline{\psi }(x)\partial_\mu \psi(x)
-\left(\partial_\mu\overline{\psi}(x)\right) \psi(x)\right].
\]
Expression (44) does not depend on the mass parameter $M_1$ and includes as the usual Dirac current (at $b=0$) as well as Barut's convective terms. One can verify with the help of Eq.(1) (at $M_0=iM_1$) that the current (44) is conserved: $\partial_\mu j_\mu (x)=0$. The charge density follows from Eq.(44) and is $j_0=-ij_4=\overline{\Psi }(x)\Gamma_4 \Psi(x)$ which vanishes for neutral particles.
The canonical energy-momentum tensor, obtained from the standard procedure \cite{Bogolyubov}, is given as follows:
\begin{equation}
T_{\mu\nu}=\frac{1}{2}\left(\partial_\nu \overline{\Psi}
(x)\right)\Gamma_\mu \Psi (x)-\frac{1}{2} \overline{\Psi}
(x)\Gamma_\mu \partial_\nu\Psi (x).
\label{45}
\end{equation}
We have taken into account here that Lagrangian (42) vanishes for fields obeying the field equations (13) (for $M_0=iM_1$) and (21). From Eqs.(9),(11),(45), one finds the energy-momentum tensor
\[
T_{\mu\nu}=\frac{1}{2}\overline{\psi} (x)\gamma_\mu\left(1+b\gamma_5\right)
\partial_\nu\psi (x)-\frac{1}{2}\left(\partial_\nu\overline{\psi} (x)\right)\gamma_\mu\left(1+b\gamma_5\right)
\psi (x)
\]
\vspace{-7mm}
\begin{equation} \label{46}
\end{equation}
\vspace{-7mm}
\[
+\frac{a}{2M}\biggl[\left(\partial_\mu\overline{\psi}
(x)\right)\partial_\nu \psi (x)-\left(\partial_\mu\partial_\nu \overline{\psi}
(x)\right)\psi (x)+ \left(\partial_\nu \overline{\psi}
(x)\right)\partial_\mu\psi (x) -\overline{\psi}
(x)\partial_\nu\partial_\mu\psi (x)\biggr] ,
\]
that is conserved: $\partial_\mu T_{\mu\nu}=0$. The canonical energy-momentum tensor (46) can be symmetrized by the Belinfante procedure. The energy density is given by ${\cal E}=T_{44}$. At $b=0$, $M_1=0$, one comes to expressions of $j_\mu$ and $T_{\mu\nu}$ obtained in \cite{Kruglov}.

\section{Non-Minimal Electromagnetic Interactions}

In this section, we consider the general case with four independent parameters,
$a$, $b$, $M$, and $M_0$. The minimal electromagnetic interaction is introduced by the
substitution $\partial _\mu \rightarrow D_\mu =\partial _\mu
-ieA_\mu $, where $A_\mu $ is the four-vector potential of the
electromagnetic field. We also consider the non-minimal electromagnetic
interaction by adding terms with two parameters $k_0$,
$k_1$ characterizing fermion anomalous electromagnetic
interactions. The first-order relativistic wave equation with minimal and non-minimal
interactions reads (see also \cite{Kruglov} for a case $b=0$, $M_0=0$)
\begin{equation}
\biggl [\Gamma_\mu D_\mu +\frac i2\left(k_0P_0+k_1
P_1\right) \Gamma_{\mu \nu }\mathcal{F}_{\mu \nu }+M+M_0\Gamma_5\biggr ]\Psi
(x)=0  ,\label{47}
\end{equation}
where $P_0=\varepsilon ^{0,0}\otimes I_4$, $P_1=\varepsilon ^{\mu
,\mu }\otimes I_4$ are the projection operators so that the relations $P_0^2=P_0$,
$P_1^2=P_1$, $P_0+P_1=1$ hold. We have introduced the matrix
\[
\Gamma_{\mu \nu }=\Gamma_\mu\Gamma_\nu -\Gamma_\nu \Gamma_\mu
\]
\begin{equation}
=\left(1-b^2\right)\varepsilon^{0,0}\otimes\left(\gamma_\mu\gamma_\nu -\gamma_\nu \gamma_\mu\right)
+a\left(\varepsilon^{\mu,\nu}-\varepsilon^{\nu,\mu}\right)\otimes I_4
\label{48}
\end{equation}
\[
+\left(\varepsilon^{\mu,0}-a\varepsilon^{0,\mu}\right)\otimes\gamma_\nu\left(1+b\gamma_5\right)
-\left(\varepsilon^{\nu,0}-a\varepsilon^{0,\nu}\right)\otimes\gamma_\mu
\left(1+b\gamma_5\right).
\]
From Eq.(47), one can obtain equations for Dirac's spinor (bispinor) $\psi$ and vector-spinor $\psi_\mu$ with the help of Eqs.(9),(11),(48):
\[
\left[ \gamma_\nu \left(1+b\gamma_5\right)D_\nu +ik_0\left(1-b^2\right) \gamma_\mu \gamma_\nu
\mathcal{F}_{\mu \nu} + M+M_0\gamma_5\right]\psi (x)
\]
\vspace{-7mm}
\begin{equation} \label{49}
\end{equation}
\vspace{-7mm}
\[
+\left[aD_\mu
+ik_0a\gamma_\nu\left(1+b\gamma_5\right)\mathcal{F}_{\nu \mu}\right]\psi_\mu (x)=0 ,
\]
\begin{equation}
\left(D _\mu +ik_1\gamma_\nu\mathcal{F}_{\mu \nu}\right) \psi
(x)+ \left(M\delta_{\mu\nu}+ik_1 a\mathcal{F}_{\mu
\nu}\right)\psi_\nu (x)=0 ,
 \label{50}
\end{equation}
where $\mathcal{F}_{\mu \nu }=\partial _\mu A_\nu -\partial _\nu
A_\mu $ is the strength of the electromagnetic field. Solving the matrix equation (50) for
vector-spinor $\psi_\mu$ and replacing it in Eq.(49), we can find an equation for bispinor $\psi$.
Such equation includes as minimal as well as non-minimal electromagnetic interactions of fermions within
our model. To clear up the physical meaning of the parameters $k_0$, $k_1$, one can consider  non-relativistic limit of Eqs.(49),(50). We leave this for further investigations.

\section{Schr\"{o}dinger Form of the Equation}

Let us consider minimal interactions of fermions by putting $k_0=k_1=0$ in Eq.(47). To obtain the Schr\"{o}dinger form of the equation and quantum-mechanical Hamiltonian we use the method of the work
\cite{Kruglov3}. Thus, we rewrite Eq.(47) as follows:
\begin{equation}
i\Gamma _4\partial _t\Psi (x)=\biggl [\Gamma_aD_a+M+M_0\Gamma_5+eA_0\Gamma _4\biggr ]\Psi (x).
\label{51}
\end{equation}
The matrix $\Gamma_4$ obeys the equation as follows (see Appendix B):
\begin{equation}
\Gamma_4^4-(1+2a-b^2)\Gamma_4^2 +a^2 \Lambda =0 ,
\label{52}
\end{equation}
where the projection operator $\Lambda$ ($\Lambda^2=\Lambda$) is given by
\begin{equation}
\Lambda =\left(\varepsilon ^{0,0 }+ \varepsilon ^{4,4}\right)\otimes I_4 .
 \label{53}
\end{equation}
The matrix $\Lambda$ extracts the $8$-dimensional sub-space of the dynamical
components of the wave function (9): $\Phi(x)=\Lambda \Psi(x)$. To separate the dynamical and non-dynamical components of the wave function $\Psi (x)$, we introduce the projection operator
$\Pi $ ($\Pi^2=\Pi$):
\begin{equation}
\Pi =1-\Lambda = \varepsilon ^{m,m }\otimes I_4 , \label{54}
\end{equation}
which defines non-dynamical components $\Omega=\Pi \Psi (x)$. Let us introduce the matrix
\begin{equation}
B=\frac{1}{a^2}\left[(1+2a-b^2)\Gamma_4-\Gamma_4^3\right],
\label{55}
\end{equation}
so that $B\Gamma_4=\Lambda$. Multiplying Eq.(51) by the matrix $B$
and taking into account Eq.(55), one obtains
\[
i\partial _t \Phi (x)=eA_0 \Phi (x)
\]
\vspace{-7mm}
\begin{equation} \label{56}
\end{equation}
\vspace{-7mm}
\[
+\frac{1}{a^2}\left[(1+2a-b^2)\Gamma_4-\Gamma_4^3\right]\left
(\Gamma_a D_a+M +M_0\Gamma_5\right)\Psi (x).
\]
From the equality $\Lambda+\Pi=1$, we find the relation
$\Psi(x)=\Phi(x)+\Omega (x)$. One can eliminate the
non-dynamical components $\Omega (x)$ from Eq.(56). Indeed, multiplying
Eq.(51) by the matrix $\Pi$, with the help of equations (see Appendix B) $\Pi \Gamma_4
=0$, $\Pi \Gamma_5=0$, $\Pi \Gamma_a\Pi=0$, we obtain
\begin{equation}
\Pi \Gamma_a D_a \Phi(x)+M\Omega (x)=0 . \label{57}
\end{equation}
Taking into account the equality $\Psi(x)=\Phi(x)+\Omega (x)$, and replacing $\Omega(x)$ from Eq.(57) into Eq.(56), one finds the Schr\"{o}dinger form of the equation
\begin{equation}
i\partial _t\Phi (x)=\mathcal{H}\Phi (x) , \label{58}
\end{equation}
where the quantum-mechanical Hamiltonian is given by
\[
\mathcal{H}=eA_0 + \frac{1}{a^2}\left[(1+2a-b^2)\Gamma_4-\Gamma_4^3\right]
\]
\vspace{-7mm}
\begin{equation} \label{59}
\end{equation}
\vspace{-7mm}
\[
\times\left(\Gamma_a D_a+M +M_0\Gamma_5\right)\left(1-\frac{1}{M}\Pi\Gamma_b D_b\right).
\]
The $8$-component wave function $\Phi(x)$ in Eq.(58) presents only dynamical components
which describe fermionic fields with two mass
states. The Hamiltonian (59) can be written in the simple form using the elements of the entire
matrix algebra (see Appendix B)
\[
\mathcal{H}= eA_0 + \frac{M}{a}\left[a \varepsilon
^{0,4}\otimes I_4 +\varepsilon ^{4,0}\otimes I_4-\varepsilon
^{4,4}\otimes \gamma_4(1+b\gamma_5)\right]
\]
\vspace{-7mm}
\begin{equation} \label{60}
\end{equation}
\vspace{-7mm}
\[
+\frac{M_0}{a}\varepsilon ^{4,0}\otimes\gamma_5
+\frac{1}{a}\left[\varepsilon ^{4,0}\otimes \gamma_m(1+b\gamma_5)\right] D_m
-\frac{1}{M}\left(\varepsilon ^{4,0}\otimes I_4\right)D_m^2 .
\]
Eq.(58) with the help of Eq.(60) takes the form of two equations
\[
i\partial _t\psi (x)= eA_0 \psi (x)+ M\psi_4 (x) ,
\]
\begin{equation}
i\partial _t\psi_4 (x)=\left[eA_0-\frac{M}{a}\gamma_4(1+b\gamma_5)\right]
\psi_4 (x)
\label{61}
\end{equation}
\[
+\frac{1}{a}\left[M+M_0\gamma_5+\gamma_m(1+b\gamma_5)
D_m-\frac{a}{M}D_m^2 \right]\psi (x) .
\]
Eqs.(61) can also be obtained from Eqs.(49),(50), at
$k_0=k_1=0$, after the exclusion of non-dynamical
(auxiliary) components $\psi_m (x)=-(1/M)D_m \psi (x)$. So, only
components with time derivatives enter Eqs.(61) and Eq.(58).
In some applications the Schr\"{o}dinger form of an equation has advantage
because it contains only dynamical components. In addition, the matrix Hamiltonian (60)
possesses only non-zero $8\times 8$ matrix block because the wave function $\Phi$ has 8 components.

\section{Field Quantization}

In this section, we consider the case $M_0=M_1=0$ but $b\neq 0$. Then there are three independent parameters $a$, $b$, $M$, and Eqs.(8) are valid. Here, we follow to the work \cite{Kruglov2} very closely. The general solution to Eqs.(13),(21) (at $M_0=M_1=0$) in the second quantized theory  can be written as follows:
\[
\Psi_\tau(x)=\sum_{s}\left[a_{\tau,s}\Psi^{(+)}_{\tau,s}(x) +
b^+_{\tau,s}\Psi^{(-)}_{\tau,s}(x)\right]
\]
\vspace{-7mm}
\begin{equation} \label{62}
\end{equation}
\vspace{-7mm}
\[
=\sum_{p,s}\frac{1}{\sqrt{2p_0 V}}\left[a_{\tau,p,s}U_{\tau,s}(
p)\exp( ipx)+ b^+_{\tau,p,s}U_{\tau,s}(- p)\exp(- ipx)\right] ,
\]
\[
\overline{\Psi}_\tau(x)=\sum_{s}\left[a^+_{\tau,s}\overline{\Psi^{(+)}_{\tau,s}}(x)
+ b_{\tau,s}\overline{\Psi^{(-)}_{\tau,s}}(x)\right]
\]
\vspace{-7mm}
\begin{equation} \label{63}
\end{equation}
\vspace{-7mm}
\[
=\sum_{p,s}\frac{1}{\sqrt{2p_0
V}}\left[a^+_{\tau,p,s}\overline{U}_{\tau,s}( p)\exp(- ipx)+
b_{\tau,p,s}\overline{U}_{\tau,s}(- p)\exp( ipx)\right] ,
\]
for the values of the quantum number $\tau=1,2$ corresponding to two
mass states (22). We introduce the creation and
annihilation operators of particles $a^+_{\tau,p,s}$, $a_{\tau,p,s}$, and antiparticles $b^+_{\tau,p,s}$, $b_{\tau,p,s}$. They obey the commutation relations
\[
\{a_{\tau,p,s},a^+_{\tau',p',s'}\}=\delta_{ss'}\delta_{\tau\tau'}
\delta_{pp'} ,~~~\{a_{\tau,p,s},a_{\tau',p',s'}\}=0
,~~~\{a^+_{\tau,p,s},a^+_{\tau',p',s'}\}=0 ,
\]
\begin{equation}
\{b_{\tau,p,s},b^+_{\tau',p',s'}\}=\delta_{ss'}\delta_{\tau\tau'}
\delta_{pp'} ,~~~\{b_{\tau,p,s},b_{\tau',p',s'}\}=0
,~~~\{b^+_{\tau,p,s},b^+_{\tau',p',s'}\}=0 , \label{64}
\end{equation}
\[
\{a_{\tau,p,s},b_{\tau',p',s'}\}=\{a_{\tau,p,s},b^+_{\tau',p',s'}\}
=\{a^+_{\tau,p,s},b_{\tau',p',s'}\}=\{a^+_{\tau,p,s},b^+_{\tau',p',s'}\}=0
.
\]
The energy density of particles and antiparticles is given by
\begin{equation}
{\cal E}= T_{44}=\frac{i}{2}\overline{\Psi}(x)\Gamma_4\partial_0\Psi (x)
 -\frac{i}{2}\left(\partial_0\overline{\Psi}(x)\right)\Gamma_4\Psi (x).
\label{65}
\end{equation}
With the help of the normalization condition (25) and Eqs.(62)-(64),
we find from Eq.(65) the energy operator
\begin{equation}
H=\int {\cal E}d^3 x=\sum_{\tau,p,s}p_0\left(a^+_{\tau,p,s}
a_{\tau,p,s}-b_{\tau,p,s} b^+_{\tau,p,s}\right) . \label{66}
\end{equation}
One can obtain from Eqs.(62)-(64) the commutation relations of
fields for different times $t$, $t'$:
\begin{equation}
\{\Psi_{\tau M(x)},\Psi_{\tau N}(x')\}=\{\overline{\Psi}_{\tau
M}(x), \overline{\Psi}_{\tau N}(x')\} =0, \label{67}
\end{equation}
\begin{equation}
\{\Psi_{\tau M}(x),\overline{\Psi}_{\tau N}(x')\}=K_{\tau
MN}(x,x'), \label{68}
\end{equation}
\[
K_{\tau MN}(x,x')=K^+_{\tau MN}(x,x')+K^-_{\tau MN}(x,x') ,
\]
\begin{equation}
K^+_{\tau MN}(x,x')=\sum_{s}\left(\Psi^{(+)}_{\tau,s}(x)\right)_M
\left(\overline{\Psi^{(+)}_{\tau,s}}(x')\right)_N  ,\label{69}
\end{equation}
\[
K^-_{\tau MN}(x,x')=\sum_{s}\left(\Psi^{(-)}_{\tau,s}(x)\right)_M
\left(\overline{\Psi^{(-)}_{\tau,s}}(x')\right)_N .
\]
We find from Eqs.(62),(63) the functions \cite{Ahieser}
\[
K^+_{\tau MN}(x,x')=\sum_{p,s}\frac{1}{2p_0 V}\left(U_{\tau,s}(
p)\right)_M\left(\overline{U}_{\tau,s}(p)\right)_N\exp [ip(x-x')]
,
\]
\vspace{-7mm}
\begin{equation} \label{70}
\end{equation}
\vspace{-7mm}
\[
K^-_{\tau MN}(x,x')=\sum_{p,s}\frac{1}{2p_0 V}\left(U_{\tau,s}(
-p)\right)_M\left(\overline{U}_{\tau,s}(-p)\right)_N\exp
[-ip(x-x')] .
\]
With the help of Eqs.(39),(41), and taking into account the relation
$p^2=-m_\tau^2$, one finds from Eqs.(70):
\[
K^+_{\tau MN}(x)=\sum_{p}\left( \frac{i\check{p}\left(M-
i\check{p}\right)
\left(\check{p}^2+Cm_\tau^2-M^2\right)m_\tau^2}{4p_0VM^3
\left(2M^2-Cm_\tau^2\right)}\right)_{MN}\exp (ipx)
\]
\vspace{-7mm}
\begin{equation} \label{71}
\end{equation}
\vspace{-7mm}
\[
=\left( \frac{\Gamma_\mu \partial_\mu\left(M-\Gamma_\nu
\partial_\nu\right)
\left[Cm_\tau^2-M^2-(\Gamma_\mu
\partial_\mu)^2\right]m_\tau^2}{2M^3
\left(2M^2-Cm_\tau^2\right)}\right)_{MN}
\sum_{p}\frac{1}{2p_0V}\exp(ipx) ,
\]
\[
K^-_{\tau MN}(x)=\sum_{p}\left( \frac{i\check{p}\left(M+
i\check{p}\right)
\left(\check{p}^2+Cm_\tau^2-M^2\right)m_\tau^2}{4p_0VM^3
\left(2M^2-Cm_\tau^2\right)}\right)_{MN}\exp
(-ipx)
\]
\vspace{-7mm}
\begin{equation} \label{72}
\end{equation}
\vspace{-7mm}
\[
=-\left( \frac{\Gamma_\mu \partial_\mu\left(M-\Gamma_\nu
\partial_\nu\right)
\left[Cm_\tau^2-M^2-(\Gamma_\mu
\partial_\mu)^2\right]m_\tau^2}{2M^3
\left(2M^2-Cm_\tau^2\right)}\right)_{MN}
\sum_{p}\frac{1}{2p_0V}\exp (-ipx) ,
\]
where $C=1+2a-b^2$.
Exploring the singular functions \cite{Ahieser}
\[
\Delta_+(x)=\sum_{p}\frac{1}{2p_0V}\exp
(ipx),~~~~\Delta_-(x)=\sum_{p}\frac{1}{2p_0V}\exp (-ipx),
\]
\vspace{-7mm}
\begin{equation} \label{73}
\end{equation}
\vspace{-7mm}
\[
\Delta_0 (x)=i\left(\Delta_+(x)-\Delta_-(x)\right),
\]
from Eqs.(69),(71),(72), we obtain
\begin{equation}
K_{\tau MN}(x)=-i\left( \frac{\Gamma_\mu
\partial_\mu\left(M-\Gamma_\nu
\partial_\nu\right)
\left[Cm_\tau^2-M^2-(\Gamma_\mu
\partial_\mu)^2\right]m_\tau^2}{2M^3
\left(2M^2-Cm_\tau^2\right)}\right)_{MN}\Delta_0 (x) .
 \label{74}
\end{equation}
Eqs.(67)-(74) are valid for two indexes $\tau=1,2$ corresponding to two mass states (22).
One can verify with the help of Eq. (28), that functions $K^-_{\tau}(x)$, $K^+_{\tau}(x)$ obey
the equations as follows:
\begin{equation}
\left(\Gamma_\mu\partial_\mu +M\right)K^-_{\tau}(x)=0 ,
~~~~\left(\Gamma_\mu\partial_\mu +M\right)K^+_{\tau}(x)=0.
 \label{75}
\end{equation}
The vacuum expectation of chronological pairing of the operators (the propagator) is defined by \cite{Ahieser}
\[
\Psi^a_{\tau M}(x)\overline{\Psi}^a_{\tau N}(y)=K^c_{\tau MN}(x-y)
\]
\vspace{-6mm}
\begin{equation} \label{76}
\end{equation}
\vspace{-6mm}
\[
=\theta\left(x_0 -y_0\right)K^+_{\tau MN}(x-y)-\theta\left(y_0
-x_0\right)K^-_{\tau MN}(x-y) ,
\]
with theta-function $\theta(x)$. Then using of Eqs.(71),(72), we find the vacuum expectation of chronological pairing of the operators for two values $\tau=1,2$:
\[
\Psi^a_{\tau M}(x)\overline{\Psi}^a_{\tau N}(y)
\]
\vspace{-6mm}
\begin{equation} \label{77}
\end{equation}
\vspace{-6mm}
\[
=\left( \frac{\Gamma_\mu\partial_\mu
\left(M-\Gamma_\nu\partial_\nu\right)
\left[Cm_\tau^2-M^2-(\Gamma_\mu
\partial_\mu)^2\right]m_\tau^2}{2M^3
\left(2M^2-Cm_\tau^2\right)}\right)_{MN}\Delta_c
(x-y) ,
\]
where
\begin{equation}
\Delta_c (x-y)=\theta\left(x_0
-y_0\right)\Delta_+(x-y)+\theta\left(y_0 -x_0\right)\Delta_-(x-y). \label{78}
\end{equation}
In Eq.(77), we have two propagators corresponding to $\tau =1$ and $\tau=2$ (there is no summation on index $\tau$ in Eq.(77)). Thus, equation (77) obtained generalizes the expression for the vacuum expectation of chronological pairing of the operators \cite{Kruglov2} on the case of $b\neq 0$. The
density matrix (36) and the vacuum expectation of chronological pairing of the operators (77)
can be used for calculations of different quantum-electrodynamics processes of fermions within the model considered.
The anticommutator (68) for equal times can be obtained from Eq.(74) using the properties of singular
functions \cite{Ahieser}.

\section{The Novel Tachyonic Equation}

Now, we consider a particular case of a tachyonic equation at $b=\pm 1$, $M_0=0$, $a<0$. Thus, Eq.(1) becomes
\begin{equation}
\left[\gamma_\mu\left(1\pm \gamma_5 \right) \partial_\mu -\frac{a}{M}\partial_\mu^2+M \right]\psi(x)=0.  \label{79}
\end{equation}
This equation describes tachyons with the mass squared (8) at $b=\pm 1$, $m^2=M^2/a$. In this case mass states degenerated and the field possesses only one mass state. As a result, the propagator can not be presented in the form of the two propagators difference (6), and the problem of the ghost presence can be avoided. It was shown in \cite{Hawking} that in HD theory with the degenerated mass states there are not problems with unitarity. As the mass of the tachyon depends on two parameters $M$ and $a$, one can choose for simplicity $a=-1$. Then using $b=-1$, we arrive at the novel equation
\begin{equation}
\left[\gamma_\mu\left(1- \gamma_5 \right) \partial_\mu +\frac{1}{M}\partial_\mu^2+M \right]\psi(x)=0.  \label{80}
\end{equation}
describing tachyons with the mass squared $m^2=-M^2$. Thus, the tachyonic mass in Eq.(80) is $M$. One can choose another value of $b=1$ in Eq.(79). The first term in Eq.(80) contains the projection operator $(1-\gamma_5)/2$ which extracts the left (chiral) component of the bispinor $\psi(x)$. The Lagrangian corresponding to the first-order formulation of Eq.(80) is given by Eq.(42) at $M_1=0$ with the matrices (14) with $b=-1$, $a=-1$. As was already mentioned, the Dirac first-order tachyonic equation does not admit the Lagrangian formulation and, therefore, the theory of tachyonic fermions suggested has an advantage.

For superluminal particles with the tachyonic mass $M$ the energy and momentum are given by
\begin{equation}
p_0=\frac{M}{\sqrt{v^2-1}},~~~~\textbf{p}=\frac{M\textbf{v}}{\sqrt{v^2-1}},
\label{81}
\end{equation}
so that $p_0^2=\textbf{p}^2-M^2$ and $v>1$. From Eq.(81), one obtains
\begin{equation}
v^2-1=\left(\frac{M}{p_0}\right)^2.
\label{82}
\end{equation}
Eq.(82) indicates that if the mass of a tachyon is constant the speed of a particle increases with decreasing the energy. For energies much larger than the tachyonic mass $M$, the particle speed  approaches the speed of light.
From Eq.(82), we find the relation
\begin{equation}
M=p_0\sqrt{v^2-1}\simeq p_0\sqrt{2\delta},
\label{83}
\end{equation}
where the deviation from the speed of light is $\delta =v-1\ll 1$. Eq.(83) allows us to calculate the
tachyonic mass $M$ if the deviation $\delta$ is measured.
The novel theory of tachyon particles based on Eqs.(79),(80) allowing the Lagrangian formulation is of definite theoretical interest.

\section{Conclusion}

We have analyzed the model based on HD Eq.(1) with four parameters $a$, $b$, $M$, $M_0$ that
generalizes the Barut's model \cite{Barut}, \cite{Barut1} with two
parameters $a$ and $M$. This allows us to consider different cases with superluminal propagation of fermions. Thus, when parameters $a$, $b$, $M$, $M_0$ are connected by relation (2) with $\kappa_3>\kappa_2$ fermions possess the superluminal speed. Another case of superluminal propagation of fermions is realizes when $M_0=0$, $|b|<1$ and $a<(b^2-1)/2$. We pay attention here on a special chose of the tachyon equation when $b=\pm 1$, $a=-1$. Thus at $b=-1$, $a=-1$ the novel equation (80) describes tachyons with one mass state, $m_1^2=m_2^2=-M^2$. Because mass states degenerated there is no conflict in this case with unitarity (see \cite{Hawking}).
We have demonstrated that the first-order formulation of equations obtained is convenient for finding the conserved electric current and energy-momentum tensor, Schr\"{o}dinger's form of the equation and quantization. The density matrix (36) and the vacuum expectation of chronological pairing of the operators (77) allow us to calculate different quantum processes of fermions in the framework of the model based on HD equations. We also introduced two parameters $k_0$ and $k_1$ characterizing non-minimal electromagnetic interactions of
fermions. There are some questions about the interpretation of parameters $k_0$, $k_1$ introduced that requires further investigations. The Schr\"{o}dinger form of the equation is convenient because it contains only dynamical components of the wave function. In the Schr\"{o}dinger picture the wave function possesses only $8$ components for describing fermionic field with two mass states in the first-order
formalism.

It should be noted also that if $M_0$ is a real value and fermions propagate with the superluminal speed there is a difficulty of the Lagrangian formulation of the model because the value $\overline{\Psi}\Gamma_5\Psi$ is imaginary.
The same situation occurs in the tachyonic Dirac equation when $a=0$, $b=0$, $M=0$ ($a/M=0$) in Eq.(1)
as the value $\overline{\psi}\gamma_5\psi$ is imaginary. This indicates that  the tachyonic first-order
Dirac equation is not healthy. But if $M_0=iM_1$ ($M_1$ is real), no difficulties with the Lagrangian
formalism. If four parameters are restricted by Eq.(2) and $M_0=iM_1$ there is subluminal propagation of fermions. At $M_0=0$, we still have the opportunity to consider the superluminal propagation of fermions at the case $|b|<1$, $a<(b^2-1)/2$ or if $b=\pm 1$, $a<0$ and there are no difficulties with Lagrangian
formulation.

\vspace{3mm}
\textbf{Appendix A}
\vspace{3mm}

Now, we find expressions for matrices entering the first-order wave
equation (13). From Eq.(14), one obtains
\[
\check{p}\equiv p_\nu\Gamma_\nu
=I^{(0)}\otimes\hat{p}\left(1+b\gamma_5\right)+I^{(1)}\otimes I_4,~~~~~~~~~~~~~~~~~~~~~~
~~~~~~~~~~~~~~~(A.1)
\]
where $\hat{p}\equiv p_\nu\gamma_\nu$, and
\[
I^{(0)}\equiv \varepsilon^{0,0} ,~~~~I^{(1)}\equiv
p_\nu\left(\varepsilon ^{\nu,0 }+ a\varepsilon ^{0,\nu }\right)
,~~~~I^{(2)}\equiv p_\mu p_\nu \varepsilon ^{\mu,\nu } .~~~~~~~~~~~~~~~~~~~(A.2)
\]
With the help of Eq.(11), it can be verified the equations as follow:
\[
I^{(0)2}=I^{(0)} ,~~I^{(0)}I^{(1)}+I^{(1)}I^{(0)}=I^{(1)},~~
I^{(2)2}=p^2I^{(2)},
\]
\[
I^{(0)}I^{(2)}=I^{(2)}I^{(0)}=0,~~I^{(2)}I^{(1)}=p^2I^{(1)}I^{(0)},~~I^{(1)}I^{(2)}=p^2I^{(0)}I^{(1)},
~~~~~~~(A.3)
\]
\[
I^{(2)}I^{(1)}+I^{(1)}I^{(2)}=p^2I^{(1)},~I^{(1)2}=a\left(p^2I^{(0)}+I^{(2)}\right).
\]
Using Eqs.(11),(A.3), we obtain matrices:
\[
\check{p}^2 =\left(1+a-b^2\right)p^2I^{(0)}\otimes
I_4+I^{(1)}\otimes\hat{p}\left(1+b\gamma_5\right)+aI^{(2)}\otimes I_4 ,~~~~~~~~~~~~(A.4)
\]
\[
\check{p}^3
=\left(1+2a-b^2\right)p^2I^{(0)}\otimes\hat{p}\left(1+b\gamma_5\right)
+\left(1+a-b^2\right)p^2I^{(1)}\otimes
I_4
\]
\[
+aI^{(2)}\otimes \hat{p}\left(1+b\gamma_5\right),~~~~~~~~~~~~~~~~~~~~~~~~~~~~~~~~~~~~~~~~~~~~~~~~~~~(A.5)
\]
\[
\check{p}^4=p^2\left(1+2a-b^2\right)\check{p}^2-a^2p^2\left(p^2I^{(0)}\otimes I_4+I^{(2)}\otimes I_4\right),~~~~~~~~~~~~~~~~~~~~~~~~~~~~~~~~~~~~~(A.6)
\]
\[
\check{p}^5
=\left(1+a-b^2\right)\left(1+3a-b^2\right)p^4I^{(0)}\otimes\hat{p}\left(1+b\gamma_5\right)
\]
\[
 +\left(1+3a+a^2-3ab^2-2b^2+b^4\right)p^4I^{(1)}\otimes I_4
\]
\[
+a\left(1+2a-b^2\right)p^2I^{(2)}\otimes \hat{p}\left(1+b\gamma_5\right).~~~~~~~~~~~~~(A.7)
\]
The matrices (A.1)-(A.7) obey the ``minimal" matrix equation:
\[
\check{p}^5-\left(1+2a-b^2\right)p^2\check{p}^3+a^2p^4\check{p}=0.~~~~~~~~~~~~~~~
~~~~~~~~~~~~~~~(A.8)
\]
From Eqs.(A.1)-(A.8), we obtain traces of the matrices:
tr$(\check{p})$=tr$(\check{p}^3)$=tr$(\check{p}^5)=0$,
tr$(\check{p}^2)=4p^2(1+2a-b^2)$,
tr$(\check{p}^4)=4p^4[(1+2a-b^2)^2-2a^2]$. Eq.(A.8) can be used for different calculations within the
first-order formalism.

\vspace{3mm}
\textbf{Appendix B}
\vspace{3mm}

In this Appendix, we obtain matrices which can be used in the Schr\"{o}dinger form of the wave
equation. With the help of Eqs.(11),(14), we find matrices
\[
\Gamma_4^2=\left[(1+a-b^2)\varepsilon ^{0,0}+a\varepsilon ^{4,4}\right]\otimes I_4
+\left(\varepsilon ^{4,0}+a\varepsilon ^{0,4}\right)\otimes\gamma_4\left(1+b\gamma_5\right),~~~~~~(B.1)
\]
\[
\Gamma_4^3=\left[(1+a-b^2)\varepsilon ^{0,0}+a\varepsilon ^{4,4}\right]\otimes \gamma_4\left(1+b\gamma_5\right)
\]
\[
+(1+a-b^2)\left(\varepsilon ^{4,0}+a\varepsilon ^{0,4}\right)\otimes I_4,~~~~~~~~~~(B.2)
\]
\[
\Gamma_4^4=\left\{\left[(1+a-b^2)^2+a(1-b^2)\right]\varepsilon ^{0,0}+a(1+a-b^2)\varepsilon ^{4,4}\right\}\otimes I_4
\]
\[
+(1+2a-b^2)\left(\varepsilon ^{4,0}+a\varepsilon^{0,4}\right)\otimes\gamma_4\left(1+b\gamma_5\right),~~~~~~~~~~~~~~~~~~(B.3)
\]
From Eqs.(B.1),(B.3), one obtains the equation as follows:
\[
\Gamma_4^4-(1+2a-b^2)\Gamma_4^2+a^2\left(\varepsilon ^{0,0}+\varepsilon ^{4,4}\right)=0.~~~~~~~~~~~~~~~~~~~~~~~~~~~~~(B.4)
\]
In in the Schr\"{o}dinger equation, we use matrices:
\[
B=\frac{1}{a^2}\left[(1+2a-b^2)\Gamma_4-\Gamma_4^3\right]
=\frac{1}{a}\left(\varepsilon ^{4,0}+a\varepsilon
^{0,4}\right)\otimes I_4 -\frac{1}{a}\varepsilon ^{4,4}\otimes
\gamma_4\left(1+b\gamma_5\right),~~~(B.5)
\]
\[
B\Gamma_m =\frac{1}{a}\varepsilon ^{4,0}\otimes \gamma_m\left(1+b\gamma_5\right) +
\varepsilon ^{4,m} \otimes I_4 ,~~~~B\Gamma_5=\frac{1}{a}\varepsilon ^{4,0}\otimes \gamma_5 ,~~~~~~(B.6)
\]
\[
\Pi \Gamma_m = \varepsilon ^{m,0}\otimes I_4 ,~~~~B\Gamma_mD_m\Pi\Gamma_nD_n
=\varepsilon^{4,0}\otimes I_4 D_m^2.~~~~~~~~~~(B.7)
\]
Also relations
\[
B\Pi=0,~~\Gamma_5\Pi=0,~~\varepsilon^{4,m}\Phi=0~~~~~~~~~~~~~~~~~~~(B.8)
\]
hold.

\end{document}